\documentclass[fleqn,usenatbib]{rasti}

\usepackage{newtxtext,newtxmath}

\usepackage[T1]{fontenc}

\DeclareRobustCommand{\VAN}[3]{#2}
\let\VANthebibliography\thebibliography
\def\thebibliography{\DeclareRobustCommand{\VAN}[3]{##3}\VANthebibliography}

\usepackage{graphicx}	
\usepackage{amsmath}	
\usepackage{soul} 
\usepackage{fontawesome}
\usepackage{xcolor}
\usepackage{subfigure}
\usepackage{caption} 
\usepackage{booktabs} 
\usepackage{multirow} 
\usepackage{makecell} 
\usepackage{adjustbox}
\usepackage{array}

\newcommand{\prob}{\ensuremath{{p}}}
\newcommand{\data}{\ensuremath{\boldsymbol{{d}}}}
\newcommand{\evidence}{\ensuremath{z}}

\newcommand{\orcid}[1]{\href{https://orcid.org/#1}{\includegraphics[width=10pt]{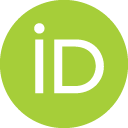}}}

\title[Bayesian model comparison for SBI]{Bayesian model comparison for simulation-based inference}

\author[A. Spurio Mancini et al.]{
	A. Spurio Mancini \orcid{0000-0001-5698-0990},$^{1}$\thanks{E-mail: a.spuriomancini@ucl.ac.uk}\thanks{Authors contributed similarly.}
	M. M. Docherty \orcid{0000-0003-2533-4077},$^{1}${$\dagger$}
	M. A. Price \orcid{0000-0002-0391-9528}$^{1}$
	and J. D. McEwen \orcid{0000-0002-5852-8890}$^{1,2}$
	\\
	$^{1}$Mullard Space Science Laboratory, University College London, Dorking, RH5 6NT, UK \\
	$^{2}$Alan Turing Institute, London, NW1 2DB, UK\\
}

\date{Accepted ---. Received ---; in original form ---}

\pubyear{2022}

\begin{document}
\label{firstpage}
\pagerange{\pageref{firstpage}--\pageref{lastpage}}
\maketitle

\begin{abstract}
	Comparison of appropriate models to describe observational data is a fundamental task of science. The Bayesian model evidence, or marginal likelihood, is a computationally challenging, yet crucial, quantity to estimate to perform Bayesian model comparison.  We introduce a methodology to compute the Bayesian model evidence in simulation-based inference (SBI) scenarios (also often called likelihood-free inference).  In particular, we leverage the recently proposed learned harmonic mean estimator and exploit the fact that it is decoupled from the method used to generate posterior samples, \textit{i.e.}\ it requires posterior samples only, which may be generated by any approach.  This flexibility, which is lacking in many alternative methods for computing the model evidence, allows us to develop SBI model comparison techniques for the three main neural density estimation approaches, including neural posterior estimation (NPE), neural likelihood estimation (NLE), and neural ratio estimation (NRE). We demonstrate and validate our SBI evidence calculation techniques on a range of inference problems, including a gravitational wave example. Moreover, we further validate the accuracy of the learned harmonic mean estimator, implemented in the \texttt{harmonic} software, in likelihood-based settings.  These results highlight the potential of \texttt{harmonic} as a sampler-agnostic method to estimate the model evidence in both likelihood-based and simulation-based scenarios.
\end{abstract}

\section{Introduction}
Bayesian model comparison provides a robust and principled statistical framework for the selection of appropriate scientific models to describe observational data.  The key quantity to perform model comparison in a Bayesian inference framework is the model evidence, or marginal likelihood, whose estimate allows one to assign relative weights to different models (see, \textit{e.g.}, \citealt{Trotta08} for a review of Bayesian model comparison, particularly in the context of cosmology). However, obtaining a precise and accurate estimate of the Bayesian model evidence is a computationally challenging task, involving a multi-dimensional integral which may quickly exceed the available computational resources for parameter spaces of even moderate dimensions.
%
A variety of techniques for computing the Bayesian model evidence have been proposed (see, \textit{e.g.}, \citealt{Friel12, llorente2020marginal} for reviews).

One of the most widely used classes of algorithms for estimating the model evidence, particularly in astrophysics and cosmology, is \textit{nested sampling} \citep{Skilling06}, a method for which posterior inferences can also be computed as a byproduct (see, \textit{e.g.}, \citealt{Buchner21, Ashton22} for reviews of nested sampling).
Popular nested sampling algorithms, such as \texttt{MultiNest} \citep{Feroz08,Feroz09} and \texttt{PolyChord} \citep{Handley15b, Handley15a} have been of remarkable success in multiple research areas.  However, some of them can struggle in high dimensional parameter spaces. The recently proposed \textit{proximal nested sampling} framework scales to very high dimensional settings \citep{Cai21}. However, proximal nested sampling is restricted to log-convex likelihoods.  Nevertheless, such likelihoods are common and so proximal nested sampling is likely to be particularly useful for inverse imaging problems.  As the name suggests, nested sampling couples the computation of the model evidence to the sampling approach, restricting its flexibility.  

The recently proposed \textit{learned harmonic mean estimator} \citep{McEwen21} for computation of the model evidence removes this restriction.  While the original harmonic mean estimator \citep{Newton94} can fail catastrophically since its variance may become very large and may not be finite \citep{neal:1994}, the learned harmonic mean solves this problem by learning an approximation of the optimal internal importance sampling target distribution \citep{McEwen21}.  Critically, the learned harmonic mean estimator requires only samples from the posterior and so is agnostic to sampling strategy, affording it great flexibility, which is crucial to the current work.

The need for efficient and reliable methods for computing the model evidence applies not only to likelihood-based settings, but also to \textit{simulation-based inference} (SBI) frameworks. In the SBI setting (sometimes referred to as \textit{likelihood-free inference}; LFI), the likelihood is either not available or too costly to be evaluated, and the inference process relies solely on the ability to simulate observables. \textit{Approximate Bayesian computation} (ABC) is the traditional, prototypical SBI technique that relies on rejection sampling of parameter sets on the basis of a similarity metric between the simulated observables and the observations (see, \textit{e.g.}, \citealt{Beaumont19}). However, ABC methods can easily require an unfeasibly large number of simulations to reach convergence, limiting their applicability.  More recently \textit{neural density estimation} techniques have been introduced to surrogate densities directly.  Such novel frameworks have seen numerous successful applications in various scientific areas, carrying great promise for the future due to their ability to avoid the evaluation of an explicit (and possibly incorrect) likelihood function, while limiting the number of simulations with clever use of cutting-edge machine learning algorithms.  For a recent review of SBI techniques we refer the reader to \citet{Cranmer20}.
Neural density estimation methods have recently found remarkable success in cosmology, with general-purpose open-source software readily available (\textit{e.g.}\ \texttt{pydelfi}\footnote{\url{https://github.com/justinalsing/pydelfi}} by \citealt{Alsing19}; \texttt{swyft}\footnote{\url{https://github.com/undark-lab/swyft}} by \citealt{Cole21}).

The development of new SBI techniques has so far mostly concentrated on optimising the task of parameter estimation, while model selection has received less attention.  Nevertheless, model selection is a critical component of a complete statistical analysis, particularly in scientific fields where selection of the appropriate model is often the fundamental question.
While there has been some consideration of model selection for SBI, the field remains nascent. \citet{brewer2011diffusive} propose a technique based on diffusive nested sampling to compute the model evidence for ABC.  However, this approach is restricted to ABC, which as discussed above can be inefficient and of limited applicability, and is not straightforwardly generalisable to modern neural density estimation approaches.

In this article we introduce a methodology to compute the model evidence, in order to facilitate Bayesian model comparison, for modern neural density estimation approaches to SBI.  Our methodology leverages the learned harmonic mean estimator \citep{McEwen21}.  We exploit the fact that the learned harmonic mean estimator is agnostic to sampling strategy and only requires samples from the posterior distribution.  In some neural density estimation approaches samples can be generated directly (\textit{e.g.}\ by pushing samples from a simple base distribution, such as a Gaussian, forward through a normalising flow; \citealt{papamakarios_normalizing_2021}).  To support such neural density estimation approaches it is therefore essential that model evidence computation is decoupled from sampling strategy, as is the case with the learned harmonic mean estimator.

The remainder of this article is structured as follows. In Sec.~\ref{sec:bayesian_model_comparison} and Sec.~\ref{sec:sbi_parameter_estimation} we concisely review, respectively, Bayesian model comparison and neural density estimation approaches to SBI.  In Sec.~\ref{sec:sbi_model_comparison} we introduce our methodology to perform Bayesian model comparison in the context of SBI, leveraging the learned harmonic mean estimator \citep{McEwen21}.  We present algorithms to compute the Bayesian model evidence for each of the three main neural density estimation approaches to SBI that are reviewed by \citet{Cranmer20}. In Sec.~\ref{sec:numerical_experiments} we report the results from numerical experiments that demonstrate and validate our methodology. Finally, we conclude in Sec.~\ref{sec:conclusions} with a review of our main findings.

\section{Bayesian model comparison}\label{sec:bayesian_model_comparison}
We review here the fundamentals of Bayesian model comparison, focusing on the challenges associated with estimation of the model evidence. For a more extensive review we refer the reader to, \textit{e.g.}, \citet{Trotta08}. We also summarise the key concepts underlying the learned harmonic mean estimator since it is an integral component of the current work (we refer the reader to \citealt{McEwen21} for further details).

\subsection{Bayesian model evidence}
The definition of model evidence in a Bayesian statistical framework follows directly from Bayes' theorem. For a given model $\mathcal{M}$, Bayes' theorem describes the connection between the conditional probabilities of model parameters $\boldsymbol{\theta}$ and data $\data$:
\begin{align}\label{eq:bayes}
	\prob ( \boldsymbol{\theta} \vert \data, \mathcal{M} )
	= \frac{\prob ( \data \vert \boldsymbol{\theta}, \mathcal{M} ) \prob ( \boldsymbol{\mathrm{\theta}} \vert \mathcal{M} )}{\prob ( \data \vert \mathcal{M} )} ,
\end{align}
where
$\prob ( \boldsymbol{\theta} \vert \data, \mathcal{M} )$ is the posterior distribution of the parameters, given the observed data $\data$ and the assumed model $\mathcal{M}$,
$\prob ( \data \vert \boldsymbol{\theta}, \mathcal{M} ) $ is the likelihood function of the data $\data$ given parameters $\boldsymbol{\mathrm{\theta}}$ and model $\mathcal{M}$,
$\prob ( \boldsymbol{\mathrm{\theta}} \vert \mathcal{M} )$ is the prior distribution of model parameters $\boldsymbol{\theta}$ for a given model $\mathcal{M}$,
and
$\prob ( \data \vert \mathcal{M} )$ is the model evidence, \textit{i.e.}\ the probability of data $\data$ for a given model $\mathcal{M}$.
The Bayesian model evidence is given by the normalisation factor of the posterior distribution $\prob ( \boldsymbol{\theta} \vert \mathcal{M}, \data )$:
\begin{align}\label{eq:evidence}
	\evidence = \prob ( \data \vert \mathcal{M} ) = \int \mathrm{d} \boldsymbol{\theta} \, \prob ( \data \vert \boldsymbol{\theta}, \mathcal{M} ) \prob ( \boldsymbol{\mathrm{\theta}} \vert \mathcal{M} ).
\end{align}

Since the model evidence is a normalisation factor for the posterior distribution and is independent of the model parameters, the evidence is usually disregarded in parameter estimation tasks. However, for model selection the evidence becomes the crucial quantity to compute. Being the integral of the likelihood over the prior (\textit{cf.}\ Eq.~\ref{eq:evidence}), the evidence allows one to assign relative weights to different models. The evidence ratio between two competing models $\mathcal{M}_1$ and $\mathcal{M}_2$ enters the expression for the comparison of their posterior distributions, which again follows from Bayes' theorem:
\begin{align}
	\frac{\prob ( \mathcal{M}_1 \vert \data )}{\prob ( \mathcal{M}_2 \vert \data )} = \frac{\prob ( \data  \vert \mathcal{M}_1 ) \prob ( \mathcal{M}_1 )}{\prob ( \data  \vert \mathcal{M}_2 ) \prob ( \mathcal{M}_2 )}.
\end{align}
In many cases \textit{a priori} probabilities $\prob (\mathcal{M}_1)$ and $\prob (\mathcal{M}_2)$ of the two models are considered to be equal, hence the ratio of posterior distributions becomes equivalent to the evidence ratio or Bayes factor
\begin{align}\label{eq:bayes_factor}
	B_{12} = \frac{\prob ( \data  \vert \mathcal{M}_1 )}{\prob ( \data  \vert \mathcal{M}_2 )} = \frac{\evidence_1}{\evidence_2}.
\end{align}
For notational brevity, henceforth we drop the explicit conditioning on models unless there are multiple models under consideration.

\subsection{Algorithms for evidence estimation}
Computing the evidence for a given model is numerically challenging due to the multi-dimensional integral in Eq.~\ref{eq:evidence}. Many techniques have been proposed to tackle this challenge, such as thermodynamic integration \citep[\textit{e.g.}][]{beltran2005bayesian, Gregory05,bridges2006bayesian}, the Savage-Dickey density ratio \citep[\textit{e.g.}][]{Trotta07}, methods based on $k$-th nearest-neighbour distances in parameter space \citep{Heavens17}, nested sampling \citep{Skilling06}, and others \citep[see, \textit{e.g.},][]{Friel12, llorente2020marginal}.

Nested sampling reduces the computation of the evidence to the evaluation of a one-dimensional integral, and as a byproduct provides samples that can be used to compute posterior inferences, thus supporting both parameter estimation and model selection. Multimodal nested sampling, implemented in the \mbox{\texttt{MultiNest}}\footnote{\url{https://github.com/farhanferoz/MultiNest}} software (\citealt{Feroz08, Feroz09}; see also \citealt{Buchner14} for the Python wrapper \texttt{PyMultiNest}\footnote{\url{https://github.com/JohannesBuchner/PyMultiNest}}), has seen enormous success, with widespread application across multiple research fields, as has the slice sampling nested sampling algorithm implemented in the \texttt{PolyChord}\footnote{\url{https://github.com/PolyChord/PolyChordLite}} software~\citep{Handley15b, Handley15a}.
Proximal nested sampling \citep{Cai21} is implemented in the \texttt{ProxNest}\footnote{\url{https://github.com/astro-informatics/proxnest}} software, which has only been released very recently but is likely to be particularly useful for inverse imaging problems.

Measuring the model evidence is a numerical process that, if repeated multiple times, produces a distribution of values. Ideally, these distributions would be narrow, and even more importantly they should provide unbiased estimates of the model evidence. However, this might not always be the case \citep{Lemos22}.
It is, therefore, crucial to develop alternative ways to estimate the model evidence, so as to perform cross-checks on the final model selection statements. Throughout we use the terminology of an \textit{accurate} estimator as one with a low bias and a \textit{precise} estimator as one with a low variance.

\subsection{Learned harmonic mean estimator}
The \textit{original harmonic mean estimator} for computation of the Bayesian model evidence was first introduced by \citet{Newton94}. From Bayes' theorem it follows that the reciprocal evidence is related to the harmonic mean of the likelihood by
\begin{align}
	\rho
	\equiv \mathbb{E}_{\prob(\boldsymbol{\theta} \vert \data)} \biggl[ \frac{1}{\prob(\boldsymbol{d} \vert \boldsymbol{\theta} )} \biggr]
	= \int \mathrm{d} \boldsymbol{\theta} \, \frac{1}{\prob(\boldsymbol{d} \vert \boldsymbol{\theta} )} \, \prob(\boldsymbol{\theta} \vert \data)
	=
	\frac{1}{\evidence}
	.
\end{align}
This relation can be used to construct an estimator of the reciprocal evidence
\begin{align}
	\hat{\rho} = \frac{1}{N} \sum_{i=1}^{N} \frac{1}{\prob(\boldsymbol{d} \vert \boldsymbol{\theta}_i )},
	\quad \boldsymbol{\theta}_i \sim \prob(\boldsymbol{\theta} \vert \data),
\end{align}
using $N$ samples $\{\boldsymbol{\theta}_i\}_{i=1}^N$ of the posterior distribution $\prob(\boldsymbol{\theta} \vert \data)$. However, this estimator may present very large or even diverging variance \citep{neal:1994}.

\citet{Gelfand94} proposed a modification to the original harmonic mean estimator, what we call the \textit{re-targeted harmonic mean estimator}, introducing a normalised target distribution $\varphi (\boldsymbol{\theta})$ to define the modified estimator
\begin{align}\label{eq:rho_hat}
	\hat{\rho} = \frac{1}{N} \sum_{i=1}^{N} \frac{\varphi (\boldsymbol{\theta}_i)}{\prob(\boldsymbol{d} \vert \boldsymbol{\theta}_i ) \prob (\boldsymbol{\theta}_i)}, \quad \boldsymbol{\theta}_i \sim \prob(\boldsymbol{\theta} \vert \data),
\end{align}
from which the original harmonic mean estimator is recovered for $\varphi (\boldsymbol{\theta}) = \prob (\boldsymbol{\theta})$.

The original harmonic mean estimator can be interpreted as importance sampling, where the importance sampling distribution is the posterior and the target distribution is the prior.  It is therefore not surprising that the original estimator suffers poor variance properties since the prior is typically broader than the posterior, whereas importance sampling requires the sampling density to be broader than the target.  By introducing a new target $\varphi (\boldsymbol{\theta})$ this issue can be circumvented provided $\varphi (\boldsymbol{\theta})$ is narrower than the posterior.  Critically, however, the introduced target must be a normalised probability distribution.

The question remains: how does one set the target distribution?  A variety of approaches have been considered previously, however none have proved completely satisfactory.  One approach is to consider a multivariate Gaussian \citep{chib:1995}; however, such a target typically has tails that are too broad.  An alternative is to consider indicator functions \citep{robert:2009, vanhaasteren:2014}; however, for complicated posterior distributions these typically capture a small region of parameter space only and so are inefficient.

It was recognised by \citet{McEwen21} that the optimal target distribution is the normalised posterior
\begin{align}\label{eq:target}
	\varphi^{\mathrm{optimal}} (\boldsymbol{\theta})
	= \frac{\prob(\boldsymbol{d} \vert \boldsymbol{\theta} ) \prob (\boldsymbol{\theta})}{\evidence}.
\end{align}
While this exact quantity is \textit{a priori} inaccessible since it involves knowledge of the evidence $\evidence$ --- precisely the quantity we are attempting to estimate --- an approximation of $\varphi(\boldsymbol{\theta})$ can be derived from posterior samples by machine learning techniques.  This is the rationale of the \textit{learned} harmonic mean estimator of \citet{McEwen21}.  Moreover, the learned approximation of the posterior need not be highly accurate; but critically it must have narrower tails than the posterior. Strategies to learn appropriate targets with these properties are presented in \citet{McEwen21}.
In a nutshell, when learning models for $\varphi(\boldsymbol{\theta})$ a bespoke optimisation problem is considered that penalises the variance of the resulting learned harmonic mean estimator, while ensuring it is unbiased, with possible additional reguarlisation. This effectively ensures the tails of $\varphi(\boldsymbol{\theta})$ are contained within the posterior. Thus, instead of learning a general distribution that matches the posterior, a distribution that is effective for the subsequent evidence computation is learned.  For further details see \citet{McEwen21}.
Note that $\varphi(\boldsymbol{\theta})$ can be trained simply from samples of the posterior and evaluating the posterior density is not strictly necessary.  However, it is necessary to evalute the normalised density of $\varphi(\boldsymbol{\theta})$ once it is trained. The learned harmonic mean estimator is implemented in the \texttt{harmonic}\footnote{\url{https://github.com/astro-informatics/harmonic}} software.

We conclude this section by highlighting that the learned harmonic mean estimator produces estimates of the evidence purely from samples of the posterior distribution; there is no requirement on the specific method used for sampling, \textit{i.e.} \texttt{harmonic} is agnostic to the method used to generate posterior samples.  As we shall see later in this work, this is the key property of the learned harmonic mean estimator that allows it to be used in a variety of simulation-based inference scenarios.

\section{Simulation-Based Inference (SBI)}\label{sec:sbi_parameter_estimation}

We provide a brief overview of the main algorithms used for SBI (simulation-based inference), referring the reader to \citet{Cranmer20} for a more extensive review. The focus of recent SBI developments and existing literature is on parameter estimation, hence we discuss SBI in this context.  In Sec.~\ref{sec:sbi_model_comparison} we introduce methodologies to perform Bayesian model comparison in an SBI setting.

The original SBI methodology, based on ABC ({approximate Bayesian computation}) (see, \textit{e.g.}, \citealt{Beaumont19}), involves simulating realisations of the observables at each of the explored points in parameter space, and accepting or rejecting these points based on their similarity with the observed data, within a tolerance $\epsilon$. This rejection sampling method recovers an accurate representation of the underlying density distribution in the limit $\epsilon \to 0$, at which point the low simulation efficiency makes computational costs infeasibly high for inference of parameter spaces with even moderate dimensionality. More recently, {neural density estimation} techniques have been introduced to overcome this computational limitation by increasing simulation efficiency.  We focus the remainder of this article on neural density estimation approaches for SBI.

In contrast to ABC, neural density estimation leverages deep neural networks to approximate conditional probability densities and is able to speed up inference by orders of magnitude \citep{Papamakarios16}. Neural density estimation involves learning a conditional density estimator $q_{\boldsymbol{\phi}}$, parameterised by weights $\boldsymbol{\phi}$, to approximate a target distribution (either the posterior distribution, the likelihood function or a density ratio proportional to the likelihood) from a training set of $N$ pairs of (typically) prior samples and simulations $\{ \boldsymbol{\theta}_i, \data_i \}_{1=1}^{N}$. Provided the density estimator is sufficiently expressive, $q_{\boldsymbol{\phi}}$ will recover an accurate estimate of the target distribution in the limit $N \to \infty$.

Neural density estimation workflows have three main phases: simulation, training and inference. The simulation phase generates the training pairs $\{ \boldsymbol{\theta}_i, \data_i \}$ that are used in the training phase to tune the weights of the neural network $\boldsymbol{\phi}$ such that $q_{\boldsymbol{\phi}}$ approximates the target density. In the inference phase, $q_{\boldsymbol{\phi}}$ is then conditioned on a specific observation $\data_0$ and parameter inference is performed.

Single runs of the simulation and training phases amortises the training of the density estimator, allowing offline inference to be run on multiple different observations, aptly named \textit{amortised neural density estimation}. However, we are often interested in inference on a specific observation $\data_0$. For this, amortised neural density estimation tends to be inefficient as generating training pairs for the density estimator across the entire prior parameter support includes many points in parameter space with very low posterior density $\prob (\boldsymbol{\theta} \vert \data_0)$.

To rectify this simulation inefficiency one can run \textit{sequential neural density estimation}, where multiple rounds of simulation and training are run sequentially to ensure there is a greater focus on regions of high posterior density. This is done by generating simulations from an alternative prior proposal distribution $\tilde{\prob} (\boldsymbol{\theta})$. This proposal distribution is iteratively updated between rounds such that for $R$ rounds, the proposal posterior of the $i$-th round $\tilde{\prob}_{i}(\boldsymbol{\theta} \vert \data_0)$ becomes the proposal prior for the subsequent round $\tilde{\prob}_{i+1}(\boldsymbol{\theta})$. This sequential approach can further increase simulation efficiency by orders of magnitude compared to the amortised counterpart \citep{Papamakarios16}, at the expense of forgoing observation-agnostic flexibility. Truncation schemes \citep{2020arXiv201113951M, 2021ANIPS..34..129M, 2022arXiv221004815D, Karchev_2022} also follow this sequential approach, by truncating the prior distribution at each sequential step in order to reduce the total number of simulations.

We briefly review the three main approaches to neural density estimation \citep{Cranmer20} \citep[see also][for a benchmark of the various algorithms]{pmlr-v130-lueckmann21a}. When referring to variants of these implementations we follow the nomenclature of \citet{Durkan20}.

\subsection{Neural posterior estimation (NPE)}
Neural posterior estimation (NPE) was first introduced by \citet{Papamakarios16} and involves training a conditional density estimator to approximate the posterior density, such that \\ $q_{\boldsymbol{\phi}} (\boldsymbol{\theta} \vert \data) \to \prob (\boldsymbol{\theta} \vert \data)$, by minimising the loss function
\begin{align}\label{eq:npe_loss}
	\mathcal{L}({\boldsymbol{\phi}}) = \mathbb{E}_{\prob(\data \vert\boldsymbol{\theta}) \prob(\boldsymbol{\theta})}\left[ -\text{log}\,q_{\boldsymbol{\phi}}(\boldsymbol{\theta} \vert \data) \right].
\end{align}

For sequential neural posterior estimation, iteratively updating the proposal distribution between inference rounds results in the density estimator learning a proposal posterior density $\tilde\prob (\boldsymbol{\theta} \vert \data)$ that is related to the true posterior density by
\begin{align}\label{eq:prop_posterior}
	\tilde{\prob} (\boldsymbol{\theta} \vert \data) \propto  \frac{\tilde{\prob} (\boldsymbol{\theta})}{\prob (\boldsymbol{\theta})}\prob (\boldsymbol{\theta} \vert \data).
\end{align}
Three variants of NPE have been introduced to recover the true posterior from the proposal posterior.

The original neural posterior estimation method (NPE-A; \citealt{Papamakarios16}) trains a mixture density network to target the posterior distribution. A post-hoc analytical correction is then applied to the resulting proposal posterior to recover an approximation of the true posterior (\textit{cf.}\ Eq.~\ref{eq:prop_posterior}).  NPE-A considers Gaussian or Gaussian mixture proposal distributions so that the correction factor can be computed analytically.

To circumvent the requirement for analytical computation, \citet{lueckmann2017flexible} propose a method (NPE-B) where the proposal correction is embedded as an importance weight in the loss function. Whilst more flexible than NPE-A, this method has been shown to suffer poor performance as the importance weights in the loss function are susceptible to high variance, resulting in early termination of training.

Finally, \citet{Greenberg19} propose a neural posterior estimation method (NPE-C) that reparameterises the problem to recover a learned approximation $q_{\boldsymbol{\phi}} (\boldsymbol{\theta} \vert \data)$ of the true posterior from a density estimator $\tilde{q}_{\boldsymbol{\phi}} (\boldsymbol{\theta} \vert \data)$ of the proposal posterior using a tractable sum of discrete atomic proposals over the support of the true posterior. This latter approach allows more flexibility in the choice of density estimator, including cutting-edge  normalising flow models \citep{papamakarios2017masked, durkan2019neural}. In our subsequent experiments we only consider this neural posterior estimation method, which we simply refer to as NPE for the remainder of this article.

NPE learns the posterior density directly, typically for a probabilistic model from which samples can be drawn directly.  For example, samples can be drawn directly from a Gaussian mixture density network or from a normalising flow, where for the latter samples are first drawn from a simple base distribution such as a Gaussian and pushed forward through the flow to yield samples of the target distribution.  Consequently, generating samples avoids the need for Markov chain Monte Carlo (MCMC) sampling and can be performed rapidly and in parallel, significantly reducing computation time for inference.

\subsection{Neural likelihood estimation (NLE)}
Neural likelihood estimation (NLE) was first introduced by \citet{Papamakarios19} and involves training a conditional density estimator to approximate the likelihood function (considering it as a probability distribution over the data), such that $q_{\boldsymbol{\phi}} (\data \vert \boldsymbol{\theta}) \to \prob (\data \vert \boldsymbol{\theta})$, by minimising the loss function
\begin{align}
	\mathcal{L}({\boldsymbol{\phi}}) = \mathbb{E}_{\prob(\data \vert\boldsymbol{\theta}) \prob(\boldsymbol{\theta})}\left[ -\text{log}\,q_{\boldsymbol{\phi}}(\data \vert \boldsymbol{\theta}) \right].
\end{align}

In contrast to NPE, sequential NLE can be implemented without a correction between $\tilde{q}_{\boldsymbol{\phi}}(\data \vert \boldsymbol{\theta})$ and $q_{\boldsymbol{\phi}}(\data \vert \boldsymbol{\theta})$.  In principle, simulations can be generated for any proposal distribution and, consequently, simulations from all sequential rounds, not just the latest, can be used when training \citep{Papamakarios19}.

This ability to seamlessly optimise simulation efficiency, however, comes at the expense of requiring an external MCMC sampling stage to generate samples from the surrogate posterior $q_{\boldsymbol{\phi}} (\data \vert \boldsymbol{\theta}) \prob(\boldsymbol{\theta})$ for inference, which increases inference time and computational cost significantly relative to NPE, where samples can be generated directly.

\begin{figure*}
	\centering
	\includegraphics[width=\textwidth]{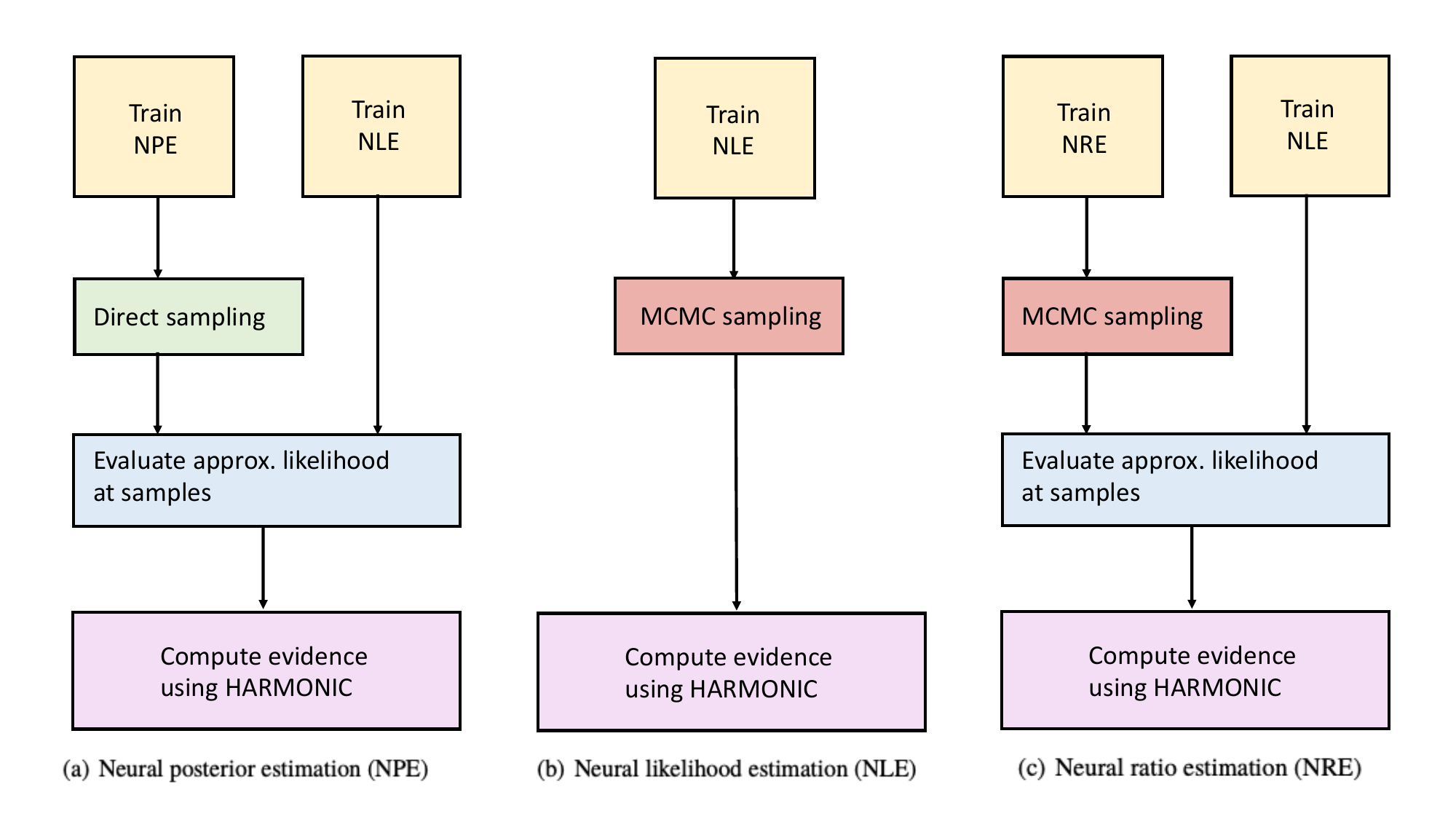}
	\caption{Schematic overview of three novel techniques that we introduce to compute the evidence in SBI settings for neural posterior, likelihood and ratio estimation methods (NPE, NLE, NRE, respectively). The \textit{yellow} training blocks represent all phases of neural density estimation, where each block can be run in an amortised or sequential setting.}
	\label{fig:schematic}
\end{figure*}

\subsection{Neural ratio estimation (NRE)}\label{sec:sbi_parameter_estimation:nre}
Neural ratio estimation (NRE) was first introduced by \citet{Hermans19} and involves approximating the posterior density $\prob (\boldsymbol{\theta} \vert \data)$ indirectly by learning a density ratio $r_{\boldsymbol{\phi}}(\data, \boldsymbol{\theta})$ that is proportional to the likelihood, where $\boldsymbol{\phi}$ denotes the model weights. This is done by training a binary classifier to discriminate samples drawn from the joint and marginal distributions of training pairs.  The classifier then learns the ratio
\begin{align}\label{eq:ratio}
	r_{\boldsymbol{\phi}}(\data, \boldsymbol{\theta}) = \frac{\prob (\data, \boldsymbol{\theta})}{\prob (\data)\prob (\boldsymbol{\theta})} =
	\frac{\prob (\data \vert \boldsymbol{\theta})}{\prob (\data)} =
	\frac{\prob (\boldsymbol{\theta} \vert \data)}{\prob (\boldsymbol{\theta})}.
\end{align}
%
A further NRE variant was devised by \citet[][]{Durkan20}. This variant extends the binary classifier to a multi-class one, hence we adopt this variant for our numerical experiments and simply refer to it as NRE for the remainder of this paper.

Similarly to NLE, an additional MCMC sampling step can be used to generate samples from the surrogate posterior $r_{\boldsymbol{\phi}}(\data, \boldsymbol{\theta}) \prob(\boldsymbol{\theta})$.  As with NLE, this increases inference time and computation cost significantly relative to NPE.  Alternatively, however, one can sample from the prior when it is tractable and incoporate approximate importance sampling weights (given by the ratio itself).

\section{Bayesian Model Comparison For SBI}\label{sec:sbi_model_comparison}

We discussed the importance and challenge of computing the model evidence for Bayesian model selection in Sec.~\ref{sec:bayesian_model_comparison}, which is a fundamental component of many scientific analyses.  Separately in Sec.~\ref{sec:sbi_parameter_estimation} we discussed three recent neural density estimation techniques for parameter estimation in an SBI (simulation-based inference) setting, which offer great promise for scientific analyses where the likelihood is often intractable or too costly to be evaluated.  In this section we unify these two topics by introducing a methodology to compute the Bayesian model evidence in all of the three neural density estimation approaches.
The evidence computation technique corresponding to each neural density estimation approach is represented schematically in Fig.~\ref{fig:schematic}. Our approaches support density estimation training phases run in either an amortised or sequential setting.

\subsection{Neural posterior estimation (NPE)}\label{sec:sbi_model_comparison:npe}

Our approach to compute the evidence for NPE is shown schematically in the left hand panel of Fig.~\ref{fig:schematic}.  We use NPE to learn an approximation $q_{\boldsymbol{\psi}}^{\text{NPE}} (\boldsymbol{\theta} \vert \data)$ of the posterior, parameterised by network weights $\boldsymbol{\psi}$.  This approach provides the ability to rapidly generate samples directly from the surrogate posterior, \textit{i.e.}\ $\boldsymbol{\theta}_i \overset{\text{direct}}{\sim} q_{\boldsymbol{\psi}}^{\text{NPE}}(\boldsymbol{\theta} \vert \data)$.
While NPE also provides the ability to evaluate the surrogate normalised posterior, the normalisation constant itself, \textit{i.e.}\ the model evidence, is not accessible.  To compute the model evidence we therefore adopt the learned harmonic mean estimator, using the samples drawn directly from the surrogate posterior.  For the learned harmonic mean estimator it is also necessary to evaluate the likelihood at sample positions (see Eq.~\ref{eq:rho_hat}), hence we adopt NLE to provide a surrogate likelihood.  Using NLE we learn an approximation $q_{\boldsymbol{\phi}}^{\textrm{NLE}} (\data \vert \boldsymbol{\theta})$ of the likelihood, parameterised by a separate set of network weights $\boldsymbol{\phi}$.
With a set of posterior samples and the surrogate likelihood learned by NLE to hand, we use the learned harmonic mean estimator to obtain an estimate of the reciprocal of the model evidence by
\begin{align}\label{eq:npe_lfi}
	\hat{\rho} = \frac{1}{N} \sum_{i=1}^{N} \frac{\varphi (\boldsymbol{\theta}_i)}{q_{\boldsymbol{\phi}}^{\textrm{NLE}} (\data \vert \boldsymbol{\theta}_i) \prob (\boldsymbol{\theta}_i)}, \quad \boldsymbol{\theta}_i \overset{\text{direct}}{\sim} q_{\boldsymbol{\psi}}^{\text{NPE}} (\boldsymbol{\theta} \vert \data).
\end{align}

The proposed approach to compute the model evidence in the NPE setting does involve training two neural density estimators, both NPE and NLE.  However, it does not require any MCMC sampling.  Samples can be generated directly from the surrogate posterior learned by NPE (\textit{e.g.}\ by pushing samples from a simple base distribution such as a Gaussian through a normalising flow), which is highly efficient and can also be computed in parallel.

Given trained NPE and NLE surrogate densities, an alternative na\"ive technique can also be considered to estimate the model evidence.  For any model parameter $\boldsymbol{\theta}$ the ratio of the unnormalised surrogate posterior, formed from the surrogate likelihood and prior, to the normalised surrogate posterior, \textit{i.e.}
\begin{equation}
	\frac{q_{\boldsymbol{\phi}}^{\textrm{NLE}} (\data \vert \boldsymbol{\theta}) \prob (\boldsymbol{\theta})}
	{q_{\boldsymbol{\psi}}^{\text{NPE}} (\boldsymbol{\theta} \vert \data)}
	,
\end{equation}
provides an estimate of the evidence.  An estimate of the evidence is thus recovered for a single parameter $\boldsymbol{\theta}$, which need not be drawn from any particular distribution.  However, such an estimate of the evidence will be incredibly noisy, \textit{i.e.}\ will have an extremely large variance.  Many parameters can be used to generate many estimates of the evidence that can be averaged.  Nevertheless, the resulting estimate of the evidence remains highly noisy with a very large variance. This na\"ive estimator relies on the ratio of two approximate quantities, hence approximation errors compound.  Contrast this with the learned harmonic mean estimator.  While our learned harmonic mean approach does use NPE to learn a surrogate posterior $q_{\boldsymbol{\psi}}^{\text{NPE}} (\boldsymbol{\theta} \vert \data)$, the density is never evaluated.  We only require samples from the corresponding distribution.  The learned harmonic mean does require learning the importance target $\varphi(\boldsymbol{\theta})$, and this is indeed learned to approximate the posterior, but the target need only be normalised and have tighter tails than the true posterior $\prob (\boldsymbol{\theta} \vert \data)$ --- it does not need to be an accurate approximation of the posterior.  Consequently, our proposed approach to compute the evidence in the NPE setting, based on the learned harmonic mean estimator, does not suffer compounding sources of error and thus provides increased stability over the na\"ive approach.

While the focus of the current article is SBI, we also comment that the ideas presented here can also be applied to accelerate evidence computation for likelihood-based inference.  Crucially, throughout our approach to compute the evidence in the NPE setting, MCMC sampling is not required.  Posterior samples can be generated directly, rapidly and in parallel.  If a likelihood is available this can simply be substituted for the surrogate likelihood learned by NLE.  Therefore in the likelihood-based setting the approach can be altered to leverage the speed of posterior sample generation of NPE, while adopting the analytical likelihood function, to obtain a rapid estimate of the reciprocal evidence without any further computation, as described by
\begin{align}\label{eq:npe_anl}
	\hat{\rho} = \frac{1}{N} \sum_{i=1}^{N} \frac{\varphi (\boldsymbol{\theta}_i)}{\prob(\data \vert \boldsymbol{\theta}_i) \prob (\boldsymbol{\theta}_i)}, \quad \boldsymbol{\theta}_i \overset{\text{direct}}{\sim} q_{\boldsymbol{\psi}}^{\textrm{NPE}} (\boldsymbol{\theta} \vert \data).
\end{align}
Clearly in this setting NLE need not be performed.

\subsection{Neural likelihood estimation (NLE)}

Our approach to compute the evidence for NLE is shown schematically in the central panel of Fig.~\ref{fig:schematic}. We use NLE to learn an approximation of the likelihood function $q_{\boldsymbol{\phi}}^{\textrm{NLE}} (\data \vert \boldsymbol{\theta})$, parameterised by network weights $\boldsymbol{\phi}$. As is typical for NLE, this approach requires MCMC sampling to generate samples from the unnormalised surrogate posterior, \textit{i.e.}\ $\boldsymbol{\theta}_i \overset{\text{MCMC}}{\sim} q_{\boldsymbol{\phi}}^{\textrm{NLE}} (\data \vert \boldsymbol{\theta})\prob(\boldsymbol{\theta})$.
NLE also provides the ability to evaluate the surrogate likelihood.  With a set of posterior samples and the surrogate likelihood learned by NLE to hand, we use the learned harmonic mean estimator to compute an estimate of the reciprocal of the model evidence by
\begin{align}\label{eq:nle_lfi}
	\hat{\rho} = \frac{1}{N} \sum_{i=1}^{N} \frac{\varphi (\boldsymbol{\theta}_i)}{q_{\boldsymbol{\phi}}^{\textrm{NLE}} (\data \vert \boldsymbol{\theta}_i) \prob (\boldsymbol{\theta}_i)}, \quad\boldsymbol{\theta}_i \overset{\text{MCMC}}{\sim} q_{\boldsymbol{\phi}}^{\textrm{NLE}} (\data \vert \boldsymbol{\theta}) \prob (\boldsymbol{\theta}).
\end{align}

This proposed approach to compute the model evidence in the NLE setting involves training only one neural density estimator, which is decidedly more efficient than training two such estimators as required in the NPE and NRE settings (\textit{cf.}\ Sec.~\ref{sec:sbi_model_comparison:npe} and Sec.~\ref{sec:sbi_model_comparison:nre}). However, it does require MCMC sampling to generate samples from the unnormalised surrogate posterior which is required to compute the evidence using the learned harmonic mean estimator.

With a trained NLE surrogate density $q_{\boldsymbol{\phi}}^{\textrm{NLE}} (\data \vert \boldsymbol{\theta}_i)$ and the prior $\prob (\boldsymbol{\theta})$ to hand, alternative techniques that only require the likelihood function and prior could also be considered to compute an estimate of the evidence.

\subsection{Neural ratio estimation (NRE)}\label{sec:sbi_model_comparison:nre}

Our approach to compute the evidence for NRE is shown schematically in the right hand panel of Fig.~\ref{fig:schematic}. We use NRE to indirectly learn an approximation $r_{\boldsymbol{\psi}}^{\text{NRE}}(\data, \boldsymbol{\theta})\prob (\boldsymbol{\theta})$ of the posterior, parameterised by network weights $\boldsymbol{\psi}$. Similarly to our NPE approach, the normalisation constant of the surrogate posterior learned by NRE, \textit{i.e.}\ the model evidence, is not accessible. We therefore adopt the learned harmonic mean estimator to compute the model evidence, for which it is also necessary to evaluate the likelihood at sample positions (see Eq.~\ref{eq:rho_hat}), hence we adopt NLE to provide a surrogate likelihood.  Using NLE we learn an approximation $q_{\boldsymbol{\phi}}^{\textrm{NLE}} (\data \vert \boldsymbol{\theta})$ of the likelihood, parameterised by network weights $\boldsymbol{\phi}$.
With a set of posterior samples and the surrogate likelihood learned by NLE to hand, we use the learned harmonic mean estimator to obtain an estimate of the reciprocal of the model evidence by
\begin{align}\label{eq:nre_lfi}
	\hat{\rho} = \frac{1}{N} \sum_{i=1}^{N} \frac{\varphi (\boldsymbol{\theta}_i)}{q_{\boldsymbol{\phi}}^{\textrm{NLE}} (\data \vert \boldsymbol{\theta}_i) \prob (\boldsymbol{\theta}_i)},\quad\boldsymbol{\theta}_i \overset{\text{MCMC}}{\sim} r_{\boldsymbol{\psi}}^{\text{NRE}}(\data, \boldsymbol{\theta})\prob (\boldsymbol{\theta}).
\end{align}

The proposed approach to compute the model evidence in the NRE setting does involve training two neural density estimators, both NRE and NLE. Furthermore, external MCMC sampling is required above to generate samples from the trained NRE surrogate posterior.

Alternatively, since NRE provides access to an approximation of the normalised posterior by $r_{\boldsymbol{\psi}}^{\text{NRE}}(\data, \boldsymbol{\theta})\prob (\boldsymbol{\theta})$, via importance sampling one could instead sample from the prior to avoid the need for MCMC sampling:
\begin{align}
	\hat{\rho}
	= \frac{1}{N} \sum_{i=1}^{N}
	\frac{r^{\text{NRE}}_{\boldsymbol{\psi}} (\boldsymbol{d}, \boldsymbol{\theta}_i) \varphi (\boldsymbol{\theta}_i) }{q_{\boldsymbol{\phi}}^{\textrm{NLE}} (\data \vert \boldsymbol{\theta}_i) \prob (\boldsymbol{\theta}_i)},
	\quad\boldsymbol{\theta}_i \overset{\text{direct}}{\sim} \prob (\boldsymbol{\theta}).
\end{align}
However, the above estimator involves the ratio of two approximate quantities and so approximation errors compound. Furthermore, one could consider using the NRE approximation of the nomalised posterior for the learned harmonic mean target distribution $\varphi(\boldsymbol{\theta})$:
\begin{align}
	\hat{\rho}
	= \frac{1}{N} \sum_{i=1}^{N}
	\frac{\bigl[r^{\text{NRE}}_{\boldsymbol{\psi}} (\boldsymbol{d}, \boldsymbol{\theta}_i)\bigr]^2}{q_{\boldsymbol{\phi}}^{\textrm{NLE}} (\data \vert \boldsymbol{\theta}_i)},
	\quad
	\boldsymbol{\theta}_i \overset{\text{direct}}{\sim} \prob (\boldsymbol{\theta}).
\end{align}
However, such an estimator is also unlikely to be well-behaved since we have not explicitly ensured the tails of $\varphi(\boldsymbol{\theta})$ are narrower than the posterior and it is a ratio of two approximate quantities, one of which is squared, and so approximation errors compound.  We therefore do not consider these variants of the estimator further.

An alternative way to compute the Bayes factor $B_{12}$ between two competing models $\mathcal{M}_1$ and $\mathcal{M}_2$, \textit{i.e.}\ the ratio of model evidences Eq.~\ref{eq:bayes_factor}, in the NRE setting is to train an additional NRE model as a binary classifier to discriminate samples from the joint and marginal distribution of the two models, respectively.  The classifier then learns the ratio
\begin{align}
	r_{\boldsymbol{\psi}_{12}}(\data, \boldsymbol{\theta})
	= \frac{\prob (\data, \boldsymbol{\theta} \vert \mathcal{M}_1)}
	{\prob (\data \vert \mathcal{M}_2 )\prob (\boldsymbol{\theta} \vert \mathcal{M}_2)},
\end{align}
where $\boldsymbol{\psi}_{12}$ denotes the network weights for a model trained in such a manner.  Following similar notation, the standard neural ratio for a single model, say model $\mathcal{M}_1$, can be denoted $r_{\boldsymbol{\psi}_{11}}$. While it is not possible to estimate the evidence of a single model directly, the Bayes factor comparing the two models, which is the critical quantity for model comparison, can then be recovered by
\begin{align}
	B_{12} = \frac{ r_{\boldsymbol{\psi}_{12}}(\data, \boldsymbol{\theta})}{r_{\boldsymbol{\psi}_{11}}(\data, \boldsymbol{\theta})}\frac{\prob(\boldsymbol{\theta}\vert \mathcal{M}_2)}{\prob(\boldsymbol{\theta}\vert \mathcal{M}_1)}.
\end{align}
We understand this method is already known to the SBI community but were not able to locate any references discussing or applying it.  Since this approach does not lie within the family of methodologies introduced in the current article, which leverage the learned harmonic mean estimate to compute the model evidence from samples of the surrogate posterior distribution, we leave the analysis of this approach to further work.

\section{Numerical Experiments}\label{sec:numerical_experiments}
Here we present the results from our numerical experiments that demonstrate and validate our SBI evidence calculation techniques.  For validation purposes, for each problem, we compare the value of the evidence computed by our proposed approach (which we stress does \textit{not} include any knowledge of the likelihood) to values computed by likelihood-based approaches, either derived analytically (when possible) and/or computed numerically by likelihood-based algorithms (\textit{e.g.}\ by \texttt{harmonic}, \texttt{MultiNest} and/or \texttt{PolyChord}).  Of course in practical SBI settings, likelihood-based alternatives will not typically be available.  Nevertheless, it is useful to consider problems here where the likelihood is available so that we can validate our SBI evidence computation techniques against likelihood-based alternatives.  All of the SBI examples were implemented using the \texttt{sbi}\footnote{\url{https://github.com/mackelab/sbi}} software \citep{Tejero-Cantero20}.

\begin{figure*}
	\includegraphics[trim={2.5cm 1cm 2.5cm 1cm}, clip=true, width=\textwidth]{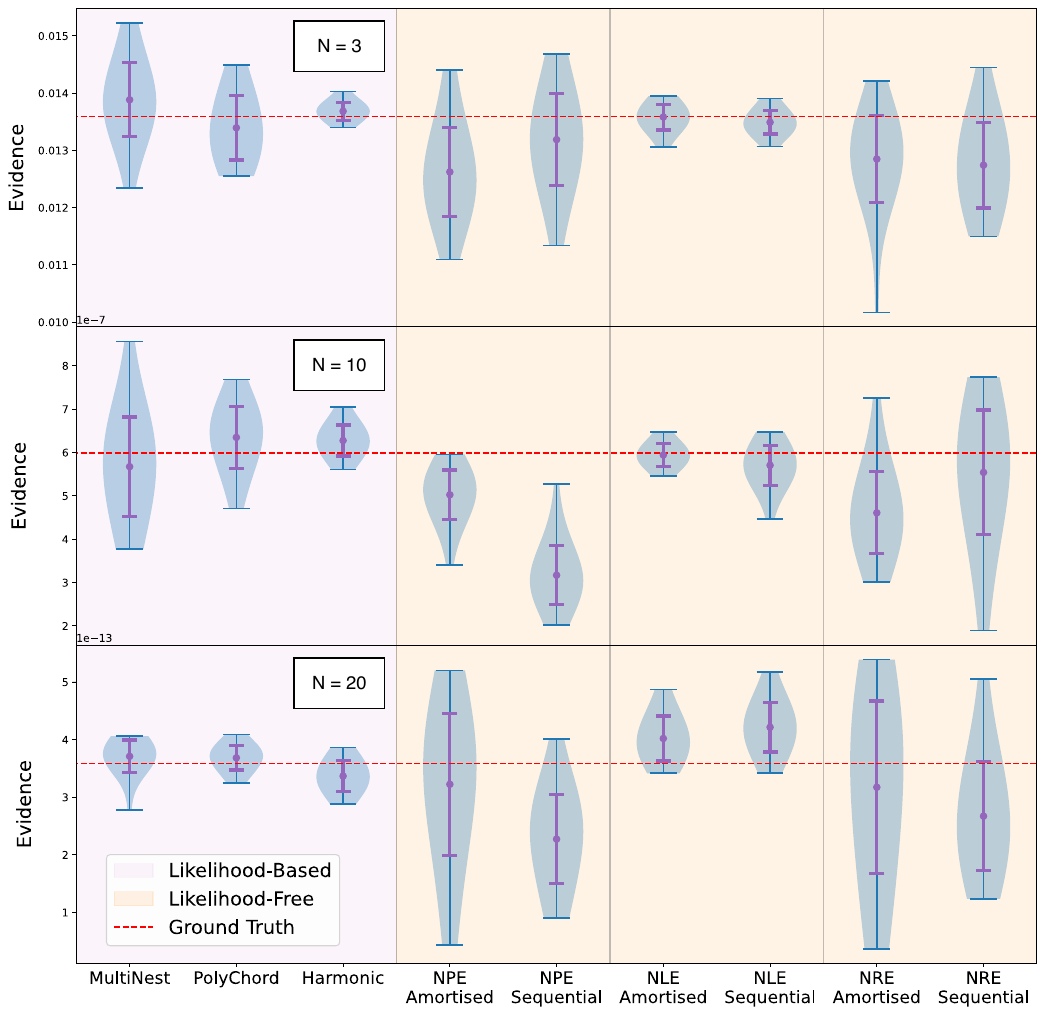}
	\caption{Model evidence values estimated with different likelihood-based and simulation-based (likelihood-free) methods for the linear Gaussian example described in Sec.~\ref{sec:linear_gaussian}, whose analytical truth value is shown by the \textit{red} dashed line. We consider three different dimensions, namely $N= \{3, 10, 20\}$, shown in the upper, central and bottom panel, respectively. For all methods we repeat the evidence estimation exercise 25 times to empirically describe the statistical distributions of the model evidence estimates, shown by the blue areas in each `violin'.  The mean and one standard deviation error bars are illustrated in purple. The \textit{pink} section of this plot shows likelihood-based results obtained with \texttt{MultiNest}, \texttt{PolyChord} and \texttt{harmonic} (the latter using samples from \texttt{emcee}). Note that the variance of the evidence estimates obtained by these three methods in the likelihood-based setting should be compared directly due to differing numbers of samples (see Sec.~{\ref{sec:linear_gaussian}} for a detailed discussion). The \textit{light brown} section of the plot shows results for the simulation-based evidence pipelines summarised in Fig.~\ref{fig:schematic}. These are all based on the use of \texttt{harmonic} to derive evidence estimates from posterior samples obtained with neural posterior estimation (NPE), neural likelihood estimation (NLE) and neural ratio estimation (NRE), in their amortised and sequential variants.}
	\label{fig:linear_gaussian}
\end{figure*}

\begin{figure*}
	\includegraphics[trim={2.5cm 0cm 2.5cm 5cm}, clip=true, width=0.9\textwidth]{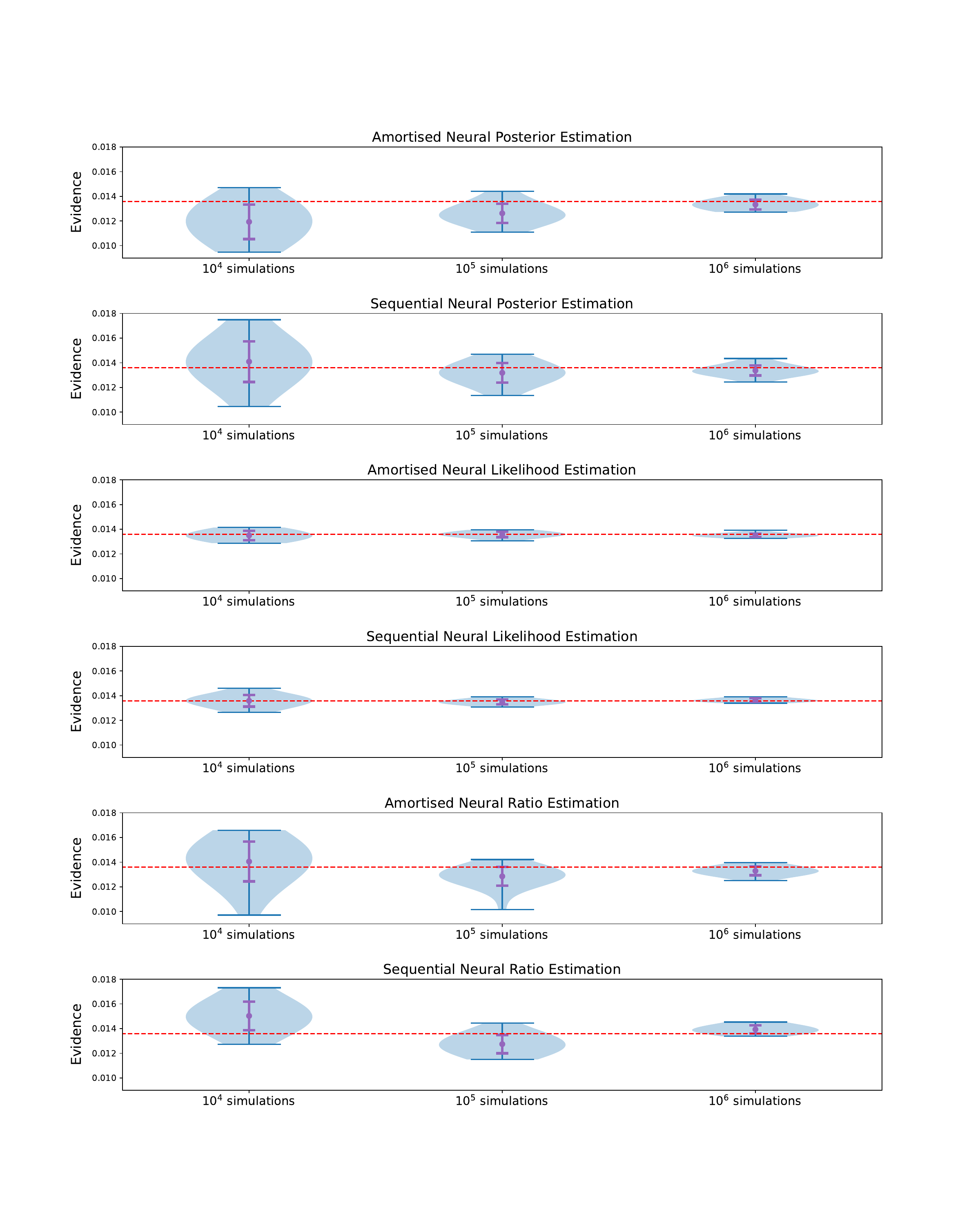}
	\vspace{-2cm}
	\caption{Dependence of the accuracy of the evidence estimates on the number of simulations used for training with simulation-based methods for the 3-dimensional linear Gaussian example, whose analytical truth value is shown by the \textit{red} dashed line.  The mean and one standard deviation error bars are illustrated in purple. We report results for the amortised and sequential versions of neural posterior estimation (NPE), neural likelihood estimation (NLE) and neural ratio estimation (NPE). The evidence pipelines are summarised in Fig.{\ref{fig:schematic}} and are all based on the use of \texttt{harmonic} to derive evidence estimates.  In all cases as the number of simulations increases the evidence estimates are less biased and the variances are reduced. These results suggests that evidence values can be computed accurately in SBI settings, although care should be taken to ensure a sufficient number of simulations are used. }
	\label{fig:number_sims}
\end{figure*}

\subsection{Linear Gaussian}\label{sec:linear_gaussian}
The first problem we consider is that of a simple simulator which linearly adds Gaussian noise $\epsilon_i$ to the value of the parameters $\theta_i$, for an arbitrary number of parameters $i = 1, \dots N$:
\begin{align}
	\mathrm{d}_i = \theta_i + \epsilon_i, \quad \epsilon_i \sim \mathcal{N}(0, 1).
\end{align}
This is a standard test problem in the \texttt{sbi} software, which we trivially generalise to arbitrary dimension $N$.
The Gaussian noise has zero mean and unit variance, and for the model parameters we assume a uniform prior $\theta_i \sim \mathcal{U}[-2, 2]$ for each component $i$. The likelihood for this model is Gaussian in the parameters $\boldsymbol{\theta} = \left\{\theta_1, \dots \theta_{N} \right\}$:
\begin{equation}
	\prob (\data \vert \boldsymbol{\theta}) = \frac{1}{(2 \pi)^{3/2}} \exp \biggl(-\frac{\bigl(\data - \boldsymbol{\theta}\bigr)^2}{2} \biggr).
\end{equation}
We assume an observation $\data_0 = ( 0, 0, 0)$. For this model the Bayesian evidence can be computed analytically:
\begin{align}\label{eq:analytical_linear_gaussian}
	\evidence & = \frac{1}{4^N \, \left( 2 \pi \right)^{N/2}} \int_{-2}^{2} \mathrm{d} \theta_1 \dots \int_{-2}^{2} \mathrm{d} \theta_N  \, \exp \biggl( -\frac{\theta_1^2 + \dots +\theta_N^2}{2} \biggr) \nonumber \\
	          & = \frac{\bigl[\text{erf}\bigl( \sqrt{2} \bigr)\bigr]^N}{4^N}.
\end{align}

\begin{figure*}
	\includegraphics[trim={2.5cm 0cm 2.5cm 2cm}, clip=true, width=0.9\textwidth]{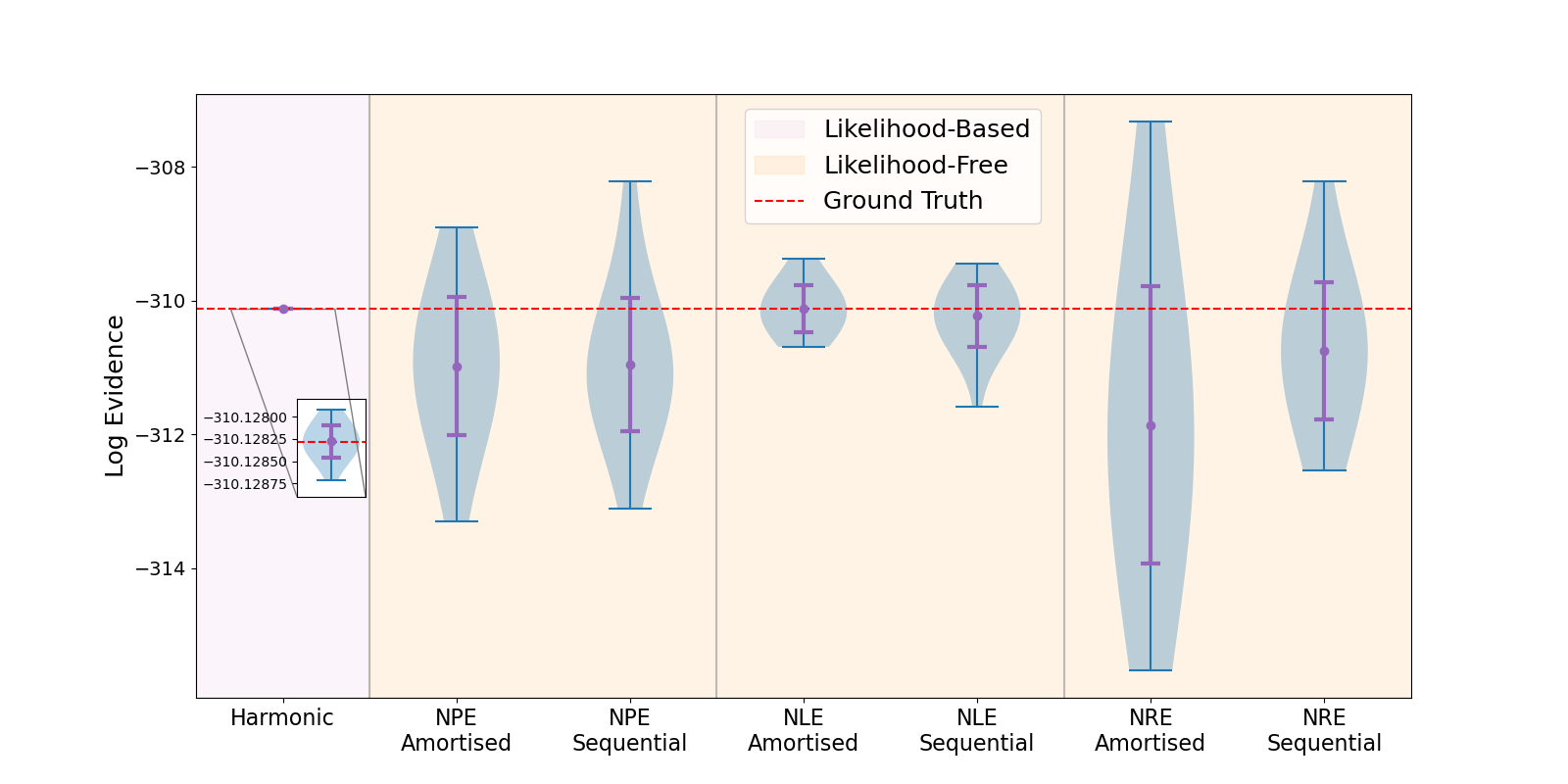}
	\caption{Model evidence values estimated with different likelihood-based and simulation-based (likelihood-free) methods for the Radiata Pine example described in Sec.~\ref{sec:radiata_pine}, whose analytical truth value is shown in \textit{red}. Colour codes and labels are consistent with Fig.~\ref{fig:linear_gaussian}.}
	\label{fig:radiata_pine}
\end{figure*}

Fig.~\ref{fig:linear_gaussian} summarises the results obtained from our model evidence estimates for the linear Gaussian problem. We consider the three cases $N=\{3, 10, 20\}$, shown in the upper, middle and bottom panels, respectively, to investigate any dependence of our evidence estimates on the number of parameters considered. In all panels, the \textit{pink} background section shows results for likelihood-based methods for validation purposes, while results for SBI (likelihood-free) methods are shown in the \textit{light brown} region.
For all methods we repeat the evidence estimation exercise 25 times to empirically describe the statistical distributions of the model evidence estimates, shown by the blue areas in each `violin' of Fig.~\ref{fig:linear_gaussian}. All of the evidence values reported in Fig.~\ref{fig:linear_gaussian} can be compared with the analytical value of Eq.~\ref{eq:analytical_linear_gaussian}, overplotted by the \textit{red} dashed line.

Likelihood-based approaches include: (a) \texttt{MultiNest}, which produces samples and evidence estimates, run with importance sampling \citep{Feroz19}, 1000 live points, efficiency sampling of 0.3 and evidence tolerance of 0.01, resulting in $\sim 10^4$ samples; (b) \texttt{PolyChord}, which also produces samples as well as evidence estimates, run using 1000 live points, resulting in $\sim 10^4$ samples; (c) \texttt{harmonic}, which produces evidence estimates from posterior samples, thus we adopt the affine sampler of \citet{Goodman10}, implemented in the  \texttt{emcee}\footnote{\url{https://github.com/dfm/emcee}} software \citep{ForemanMackey13}, to generate $10^5$ post burn-in posterior samples from 100 random walkers and adopt a hypersphere model for the learned importance target $\varphi(\boldsymbol{\theta})$, with radius equal to the square root of the number of parameters. The number of samples generated by \texttt{MultiNest} and \texttt{PolyChord} is dynamic, depending on, e.g., the tolerance parameter, whereas for \texttt{emcee} sampling for \texttt{harmonic} we adopt a conservative, fixed number of samples, since this is the configuration we will also use when considering the SBI scenarios.  Consequently, the variances of the nested sampling approaches should not be compared directly to those of the learned harmonic mean due to the differing number of samples.

We analyse the performance of the estimators in terms of their bias and variance (also adopting the terminology \textit{accuracy} and \textit{precision}, respectively, as common in the astrophysical literature).  All of the three likelihood-based methods provide unbiased average estimates of the evidence. As mentioned, the variance of these estimators should not be directly compared due to the differences in the number of samples (an analysis of the performance of nested sampling and learned harmonic mean approaches would be interesting but here we are focused on validating our proposed techniques to compute the evidence for SBI scenarios).

The results for the SBI methods are shown in the \textit{light brown} section of the plot. We report results for NPE, NLE and NRE, and for each of them we provide an estimate using the amortised as well as the sequential approach. For all methods we use $10^5$ simulations in the amortised approach, while in the sequential one we use 10 rounds with $10^{4}$ simulations each, thus totalling the same number of simulations for the two approaches. For each SBI method after training a density estimator (in an amortised or sequential fashion) we collect a total of $10^5$ posterior samples, either directly for NPE or by MCMC sampling using 100 \texttt{emcee} random walkers for NLE and NRE. We train the \texttt{harmonic} importance target model using 20\% of the samples and use the remaining 80\% to compute the evidence. As explained in Sec.~\ref{sec:sbi_model_comparison:npe} and Sec.~\ref{sec:sbi_model_comparison:nre}, NPE and NRE require an additional training of an NLE estimator to provide a surrogate likelihood.

All of the SBI evidence computation techniques provide estimates of the evidence whose distribution captures the true analytic evidence in the $N=3$ and $N=20$ cases, although there is a residual bias in many cases.  Moreover, for $N=10$ the NPE estimator distribution does not always capture the true evidence.  The fact that the estimates for dimension $N=20$ --- the more challenging setting --- do encompass the true evidence suggests that these biases may be due to insufficient training of the underlyling SBI models and/or their difficulty to scale to higher dimensions.  The variances of the SBI estimates are generally larger than the likelihood-based approaches, which is to be expected since in contrast to the likelihood-based setting we do not include any knowledge of the likelihood. We also note that each evidence estimate with an SBI method does include some training noise due to the fact that we repeat the training at every iteration.  The NLE approaches exhibit in general less bias than the NPE and NRE estimators. This may be due to the fact that the NLE approach requires only a single neural density estimator (whereas the NPE and NRE approaches require two), resulting in fewer sources of approximation error.

In Fig.~{\ref{fig:number_sims}} we investigate the dependence of the evidence estimate on the total number of simulations used in training the neural density models. We consider the case $N=3$ and we report the results for the amortised and sequential approach to NPE, NLE and NRE, varying the number of simulations from $10^4$, to $10^5$, to $10^6$. We observe that in all cases that as the number of simulations increases the evidence estimates are less biased and the variances are reduced.  This initial analysis suggests that evidence values can be computed accurately in SBI settings, although care should be taken to ensure a sufficient number of simulations are used.  A more extensive analysis of the accuracy and precision of SBI approaches for evidence calculation would be welcome, along the lines of the extensive and informative study performed by {\citet{Hermans21}}.  However, such an analysis requires substantial computational resources and is beyond the scope of the current article, where we introduce these new methodologies.

\begin{figure*}
	\includegraphics[width=.9\textwidth]{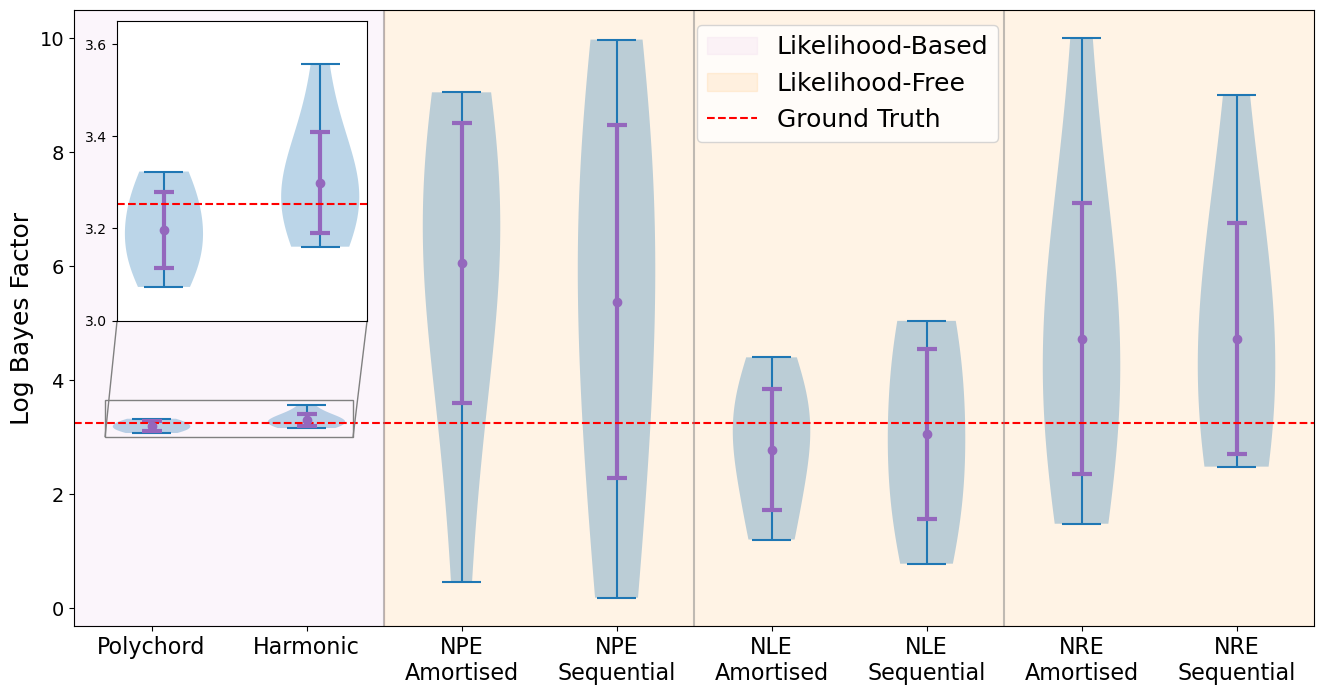}
	\caption{Logarithm of the Bayes factors between source and alternative waveform models estimated with different likelihood-based and simulation-based (likelihood-free) methods for the gravitational wave example described in Sec.~\ref{sec:gw}. Shown in \textit{red}, an estimate of the true value of the logarithm of the Bayes factor was obtained using numerical integration. Colour codes and labels are consistent with Fig.~\ref{fig:linear_gaussian} and Fig.~\ref{fig:radiata_pine}.}
	\label{fig:gw}
\end{figure*}

\subsection{Radiata pine}\label{sec:radiata_pine}

The second problem we considered is one of the classic benchmark examples used to evaluate techniques to estimate the model evidence \citep{Friel12, Williams59}. We refer to \citet{McEwen21}, who demonstrated the effectiveness of the learned harmonic mean estimator for this example, for a more detailed presentation of the problem: here we simply report the main points relevant to our evidence estimation task.

Our dataset is comprised of measurements of Radiata pine trees of the maximum compression strength parallel to the grain ${y}_i$, for $i = 1 \dots 42$. The original scientific problem can be stated in terms of two models: in Model 1 the density ${x}_i$ is assumed as a predictor for ${y}_i$, while in Model 2 the predictor is assumed to be the resin-adjusted density ${z}_i$. Both predictors are modelled with a Gaussian linear regression model, for which the value of the evidence can be derived analytically for each model. For brevity, we report results only for Model 1; calculations for Model 2 are identical (we did also experiment with this second model, finding excellent agreement with the analytical estimates of the evidence). In Model 1, denoting with $\bar{{x}} = \frac{1}{n}\sum_{i=1}^{42} {x}_i$ the average density across the trees specimens, the maximum compression strength ${y}_i$ is given by
\begin{align}
	y_i = \alpha + \beta ({x}_i - \bar{{x}}) + \epsilon_i, \quad \epsilon_i \sim \mathcal{N}(0, \tau^{-1}).
\end{align}
The model parameters are
$\{ \alpha, \beta, \tau \}$, whose prior distributions are
\begin{align}
	\alpha \sim \mathcal{N} ( \mu_\alpha,  ( r_0 \tau )^{-1} ), \;
	\beta  \sim \mathcal{N} ( \mu_\beta,  ( s_0 \tau )^{-1} ), \;
	\tau   \sim \mathrm{Ga} ( a_0, b_0 ),
\end{align}
with $( \mu_\alpha, \mu_\beta, r_0, s_0, a_0, b_0 ) = (3000, 185, 0.06, 6, 3, 2 \times 300^2)$. The evidence for this model can be computed analytically \citep[\textit{cf.}][Eq.~104]{McEwen21}; the numerical value of its logarithm is $\log \evidence = -310.12829$.

Fig.~\ref{fig:radiata_pine} summarises our findings for the Radiata pine example; the colour codes are the same as in Fig.~\ref{fig:linear_gaussian}. We repeat the same experiments as in the linear Gaussian example, except this time for simplicity we do not attempt to calculate the evidence with \texttt{MultiNest} or \texttt{PolyChord} (as this would require some effort to adapt the prior function for the Radiata pine model to be compatible with the uniform distribution on the unit cube required by these nested samplers). Therefore, for the likelihood-based case we report only numerical results obtained by applying \texttt{harmonic} to \texttt{emcee} samples, using a kernel density estimate for the learned harmonic mean estimator importance target, with radius 0.02 of the target distribution. As in the linear Gaussian example, we use 20$\%$ of $10^5$ \texttt{emcee} posterior samples from 100 random walkers to train \texttt{harmonic}, and compute an estimate of the evidence with the remaining 80$\%$. As we can see in the \textit{pink} section of the plot, this provides unbiased and tight estimates of the evidence.

In the \textit{light brown} background section of Fig.~\ref{fig:radiata_pine} we can compare results for SBI methods. The number of simulations we use to train the density estimators in the various methods is the same as in the baseline case for the linear Gaussian example, as is the number of posterior samples used to train \texttt{harmonic} and derive evidence estimates.
All of the SBI evidence pipelines provide reasonably accurate estimates of the evidence, with distribution ranges capturing the true analytic evidence.  The NPE and NRE approaches again exhibit some bias, which is nevertheless within the spread of the distribution of values. The NLE estimates again show good agreement with the reference values.

\subsection{Gravitational waves}\label{sec:gw}

The final problem we consider is a simulated measurement of a gravitational wave (GW) signal from a single interferometer. We consider a merger between two black holes of mass $M_1 = M_2 = 20 M_{\sun}$, following a similar numerical setup to the one considered by \citet{Jeffrey20}.  We compute the noiseless time series of the strain signal using the \texttt{pycbc}\footnote{\url{https://github.com/gwastro/pycbc}} software \citep{Biwer19}. We only consider the `$+$' polarization of the detector strain $h_{+, \times}$. The duration of the signal is $\sim 0.12\text{s}$, sampled at steps of duration $\sim 488 \mu \text{s}$ each. We rescale the signal by a multiplicative factor $10^{21}$ in order to work with values $\mathcal{O}(1)$. For each point in parameter space, our simulated observable is a noisy gravitational waveform obtained by adding Gaussian noise with zero mean and standard deviation $\sigma = 0.3$ to the noiseless template from \texttt{pycbc}.

For this inference problem we vary the two black hole masses $M_1, M_2$, each one over a uniform prior $\mathcal{U} \left[ 10, 30 \right] M_{\sun}$. We do not consider geometrical properties of the black holes such as spin or inclination angle, similarly to \citet{Jeffrey20}. As noted by \citet{Hermans21}, who studied a similar GW SBI problem, such a simulated experimental setup requires significant computational demand. Hence, for this example we run only 10 repetitions of each inference method to empirically describe the statistical distribution of evidence estimates. These estimates mimic a realistic scenario within GW data analysis pipelines comparing model evidences for two different numerical approximant models assumed in the generation of the noiseless template waveforms. The first of these two waveform models corresponds to the actual one used to generate the simulated observation, a reduced-order effective-one-body model \citep[{SEOBNR},][]{Taracchini14}, which we refer as the \textit{source} model. The second model we consider is an inspiral-merger-ringdown phenomenological model \citep[{IMRPhenom},][]{Hannam14}, which we refer to as the \textit{alternative} model. For this configuration, we expect to find the Bayes factors comparing models to favour the source {SEOBNR} model. We verify this numerically, taking advantage of this problem's deliberately low-dimensional parameter space, which allows us to compute an estimate of the evidence for each model using direct numerical integration. We find the logarithm of the Bayes factor computed by direct numerical integration to be $\sim3.25$, favouring the source model as anticipated.

We run the inference pipeline in multiple likelihood-based contexts, namely (a) obtaining samples and evidence estimates with \texttt{PolyChord}, using the same configuration for this method as the one described in Sec.~\ref{sec:linear_gaussian}; (b) sampling the parameter space with \texttt{emcee}, collecting the same number of posterior samples as in both previous numerical examples, and using \texttt{harmonic} to obtain an estimate of the evidence, with a kernel density estimate for the importance target distribution of radius 0.002 and 0.02 for the source and alternative models, respectively. Numerical results from these likelihood-based methods are reported in the \textit{pink} background section of Fig.~\ref{fig:gw}, where we show violin plots for log-Bayes factors computed with \texttt{PolyChord} and \texttt{harmonic}. The distribution of log-Bayes factors always favours the true underlying source model and, as in previous examples, we find strong agreement between each likelihood-based approach.

Numerical results from the SBI evidence estimates are shown in the \textit{light brown} background section of Fig.~\ref{fig:gw}. The evidence pipelines we consider are the same as described in Sec.~\ref{sec:sbi_model_comparison} and for each pipeline, the number of simulations used to train the density estimators are the same as those used for the baseline linear Gaussian and Radiata pine examples in Sec.~\ref{sec:linear_gaussian} and Sec.~\ref{sec:radiata_pine}, for both amortised and sequential approaches. We also use the same number of posterior samples to train \texttt{harmonic} and derive estimates of the evidence. We notice that all SBI methods produce log-Bayes factor estimates in general agreement with the likelihood-based ones, albeit with a comparatively larger variance than presented in Fig.~\ref{fig:linear_gaussian} and Fig.~\ref{fig:radiata_pine}, due to compounding errors from the multiple evidence estimates required to obtain Bayes factors (\textit{cf.} Eq.~\ref{eq:bayes_factor}). NLE produces more unbiased estimates of the log-Bayes factors compared to NPE and NRE --- a similar trend to that observed in the linear Gaussian and Radiata pine examples (\textit{cf.}\ Fig~\ref{fig:linear_gaussian} and Fig.~\ref{fig:radiata_pine}).

\section{Conclusions}\label{sec:conclusions}
In this article we propose a novel methodology to compute the model evidence for modern neural density estimation approaches to simulation-based inference (SBI) using the learned harmonic mean estimator. Our approach leverages the property of the learned harmonic mean estimator that it is decoupled from the sampling strategy and only requires samples of the posterior. This allows us to develop SBI model comparison techniques for all three main neural density estimation approaches: neural posterior estimation (NPE), neural likelihood estimation (NLE), and neural ratio estimation (NRE).

We demonstrate and validate our SBI evidence calculation techniques on a range of inference problems, using the learned harmonic mean estimator as implemented in the \texttt{harmonic} software. We validate all SBI evidence estimator approaches, computed using \texttt{harmonic}, against those computed by likelihood-based alternatives. We find that \texttt{harmonic} produces values of the evidence that are in excellent agreement with those computed by the likelihood-based nested sampling algorithms \texttt{MultiNest} and \texttt{PolyChord}. Our results suggest that the learned harmonic mean estimator can be reliably used as an alternative to nested sampling for evidence estimation.

We also compare the different SBI evidence computation approaches that we propose and find that the NLE evidence estimation approach provides more accurate evidence estimates compared to the NPE and NRE approaches. This result is particularly encouraging for applications to cosmological scenarios, where it is very common to perform SBI inference using the \texttt{pydelfi} software, which indeed implements (sequential) NLE to sample the posterior distribution.

Overall, our methodology and proof-of-concept analysis highlight the potential of the learned harmonic mean estimator as an additional tool for Bayesian model selection in SBI settings. Future research will focus on extending the applicability of \texttt{harmonic} to larger data and parameter spaces, as well as on estimating the variance of the estimators directly, folding in both sampling variance and neural density approximate error.  Application to more realistic examples and problems will be insightful, as well as a thorough analysis of the accuracy and precision of the proposed approach; we advocate a future study similar to \citet{Hermans21} but focused on model selection.  Our hope is that this article provides a first step towards the computation of the Bayesian model evidence in SBI scenarios, in order to facilitate principled and robust Bayesian model selection.

\section*{Acknowledgements}
We thank T. Kitching and M. Zaldarriaga for useful discussions. ASM is supported by the MSSL STFC Consolidated Grant ST/W001136/1 and the Leverhulme Trust. MMD is supported by the Science and Technology Facilities Council (STFC) Centre for Doctoral Training (CDT) in Data Intensive Science (DIS) at UCL. MAP is supported by EPSRC grant EP/W007673/1. This work used computing facilities provided by the UCL Cosmoparticle Initiative and also facilities funded by the Research Capital Investment Fund (RCIF) provided by UKRI, and partially funded by the UCL Cosmoparticle Initiative and the UCL CDT in DIS.

\section*{Data Availability}

\texttt{harmonic} is freely available at \url{https://github.com/astro-informatics/harmonic}.

\bibliographystyle{rasti}
\bibliography{references}




\bsp	
\label{lastpage}
\end{document}